\documentclass[aps,prl,twocolumn,showpacs,groupedaddress]{revtex4}
\usepackage{graphicx}  
\usepackage{dcolumn}   
\usepackage{bm}        
\usepackage{amssymb}   
 
\begin{document}

\hspace{5.2in} \mbox{FERMILAB-PUB-08-279-E}

\title{Observation of $ZZ$ production in $p\bar{p}$ collisions
at $\sqrt{s}=$1.96~TeV}

%
\author{V.M.~Abazov$^{36}$}
\author{B.~Abbott$^{75}$}
\author{M.~Abolins$^{65}$}
\author{B.S.~Acharya$^{29}$}
\author{M.~Adams$^{51}$}
\author{T.~Adams$^{49}$}
\author{E.~Aguilo$^{6}$}
\author{M.~Ahsan$^{59}$}
\author{G.D.~Alexeev$^{36}$}
\author{G.~Alkhazov$^{40}$}
% ...

\affiliation{\vspace{0.1 in}(The D\O\ Collaboration)\vspace{0.1 in}}
% ...
\affiliation{$^{6}$University of Alberta, Edmonton, Alberta, Canada,
                Simon Fraser University, Burnaby, British Columbia, Canada,
                York University, Toronto, Ontario, Canada, and
                McGill University, Montreal, Quebec, Canada}
% ...
\affiliation{$^{29}$Tata Institute of Fundamental Research, Mumbai, India}
% ...
\affiliation{$^{36}$Joint Institute for Nuclear Research, Dubna, Russia}
% ...
\affiliation{$^{40}$Petersburg Nuclear Physics Institute,
                St. Petersburg, Russia}
% ...
\affiliation{$^{49}$Florida State University, Tallahassee, Florida 32306, USA}
% ...
\affiliation{$^{51}$University of Illinois at Chicago,
                Chicago, Illinois 60607, USA}
% ...
\affiliation{$^{59}$Kansas State University, Manhattan, Kansas 66506, USA}
% ...
\affiliation{$^{65}$Michigan State University,
                East Lansing, Michigan 48824, USA}
% ...
\affiliation{$^{75}$University of Oklahoma, Norman, Oklahoma 73019, USA}
% ...
\date{August 6, 2008}

\begin{abstract}
We present an observation for $ZZ \to \ell^{+}\ell^{-}\ell^{'+} \ell^{'-}$
($\ell$, $\ell^{'}$ = $e$ or $\mu$) production in $p\bar{p}$ collisions at
a center-of-mass energy of $\sqrt{s}=1.96$~TeV. Using 1.7~fb$^{-1}$
of data collected by the D0 experiment at the Fermilab Tevatron Collider, 
we observe three candidate events with an expected background of 
$0.14^{+0.03}_{-0.02}$ events. 
The significance of this observation is 5.3 standard deviations.
The combination of D0 results in this channel, as well as in 
$ZZ\to\ell^{+}\ell^{-}\nu\bar{\nu}$, yields a significance of 5.7 standard 
deviations and a combined cross section of 
$\sigma(ZZ) = 
1.60 \pm 0.63~\mathrm{(stat.)}^{+0.16}_{-0.17}~\mathrm{(syst.)}$~pb.
\end{abstract}

\maketitle 

Studies of the pair production of electroweak gauge bosons provide
an interesting test of the electroweak theory predictions~\cite{brown}. 
In contrast
with other diboson processes, $Z$ boson pair production ($ZZ$) does not
involve trilinear gauge boson interactions within the standard model (SM).
The observation of an unexpectedly high cross section could indicate
the presence of anomalous $ZZZ$ or $ZZ\gamma$ couplings~\cite{zz-theory}.
The SM prediction for the total $ZZ$ production cross section in $p\bar{p}$
collisions at the Fermilab Tevatron Collider at $\sqrt{s}=1.96$~TeV is
$\sigma(ZZ)=1.4 \pm 0.1$~pb~\cite{Campbell:1999ah}. The requirement of 
leptonic $Z$ boson decays reduces the observable cross section, making its 
measurement rather challenging. The accumulation of integrated luminosities 
in excess of 3 fb$^{-1}$ at the Fermilab Tevatron Collider and the development 
of highly optimized event selection criteria has now made possible the direct 
observation of $ZZ$ production.

Previous investigations of $ZZ$ production have been performed both at the
Fermilab Tevatron $p\bar{p}$ and the CERN $e^+e^-$ (LEP)~\cite{lep} 
Colliders. The D0 
collaboration
reported a search for $ZZ\to\ell^{+}\ell^{-}\ell^{'+}\ell^{'-}$ 
($\ell$, $\ell^{'}$ = $e$ or $\mu$)
with 1 fb$^{-1}$ of data
that provided a 95\% C.L. limit of $\sigma(ZZ)<4.4$~pb~\cite{runiia_zz}.  
The CDF collaboration 
reported a signal for $ZZ$ production with a significance of 4.4 
standard deviations from combined $ZZ\to\ell^{+}\ell^{-}\ell^{'+}\ell^{'-}$ 
and 
$ZZ\to\ell^{+}\ell^{-}\nu\bar{\nu}$ searches, and  measured a production cross 
section of
$\sigma(ZZ)=1.4^{+0.7}_{-0.6}$~pb~\cite{cdf_zz}. 

In this Letter, we present a 
search for $Z$ boson pairs where the $Z$ bosons have decayed to either
electron or muon pairs, resulting in final states consisting
of four electrons ($4e$), four muons ($4\mu$) or two muons and
two electrons ($2\mu2e$).
Data used in this analysis were collected with the D0~detector
at the Fermilab Tevatron $p\bar{p}$ Collider at
$\sqrt{s}=1.96$ TeV between June 2006 and May 2008.
The integrated luminosities~\cite{d0lumi} for the three analyzed channels
are about 1.7 fb$^{-1}$. This result is later combined
with that from an earlier similar analysis~\cite{runiia_zz} using
data collected from October 2002 to February 2006 and corresponding to an
integrated luminosity of 1 fb$^{-1}$.

The D0 detector~\cite{run2det} consists of a 
central tracking system, comprised of a silicon microstrip tracker (SMT),
and a central fiber tracker (CFT), providing coverage to pseudorapidity
$|\eta| < 3$ ~\cite{pseudo}, both located within a 
2~T superconducting solenoidal magnet. Three
liquid argon and uranium calorimeters provide coverage 
to $|\eta| < 4$. Electromagnetic objects are well
reconstructed in the regions of the central calorimeter (CC) with
coverage to $|\eta| < 1.1$ and the end calorimeters (EC) with coverage
of $1.5 < |\eta| < 3.2$. A muon system
surrounds the calorimetry, consisting of three layers of scintillators
and drift tubes and 1.8~T iron toroids, with a coverage of 
$|\eta| < 2$.

This analysis employs a 
combination of single and dielectron triggers
for the 4e channel. Similarly, single 
and dimuon triggers are used for the $4\mu$ channel.
The $2\mu2e$ channel uses a combination of all these triggers,
and additional specific electron-muon triggers.
The triggering efficiency for events with four leptons having
high transverse momentum ($p_{T})$ that satisfy all offline 
selection requirements exceeds 99\%.

For the $4e$ channel, we require four electrons with ordered
transverse energies $E_{T} >$ 30, 25, 15, and 15~GeV, respectively. 
Electrons must be 
reconstructed either in the CC region or in the
EC region, and be
isolated from other energy clusters in the calorimeter. 
Electrons in the CC are required to satisfy identification criteria 
based on a multivariate
discriminant derived from calorimeter shower shape and 
a matched track reconstructed in the SMT and CFT. Electrons in the EC are 
not required to have a matched track, but must satisfy  more
stringent shower shape requirements.
At least two electrons must be in the CC region.
With no requirement applied on the charge of the electrons at this stage to 
increase selection efficiency,
three possible $ZZ$ combinations can be formed for each $4e$
event. Events are required to have a solution for which one $ee$ combination 
has an invariant mass
$> 70$~GeV and the other $> 50$~GeV.  
Finally, events are split
into three categories, depending on the number of
electrons in the CC region. Subsamples with two electrons,
with three electrons and with four or more electrons in the CC are
denoted as $4e_{2C}$, $4e_{3C}$,
and $4e_{4C}$, respectively.
The three exclusive subsamples contain significantly different 
levels of background contamination and thus the separation of the subsamples
provides more sensitivity to the search.

For the $4\mu$ channel, each muon must satisfy 
quality criteria
based on scintillator and wire chamber information from the muon
system, and have a matched track in the central tracker. 
We require that the four most energetic muons have ordered
transverse momenta $p_T >$ 30, 25, 15, and 15~GeV, respectively.
At least three muons in the event must be isolated,
each passing a requirement of less than $2.5$~GeV of transverse energy
deposited in the calorimeter in the annulus $0.1 < \Delta R < 0.4$
centered around the muon track~\cite{cal_iso}.
Finally, the muon is required to be well reconstructed and to
originate from the primary event vertex.
Of the three possible $ZZ$ 
combinations per event that can be formed without considering muon charge
at this stage, 
a solution is required where one $\mu\mu$ combination has an 
invariant mass $> 70$~GeV 
and the other $> 50$~GeV.

For the $2\mu2e$ channel, 
the two most energetic electrons and muons 
in an event must have $E_T(p_T)>25, 15$ GeV.  All muons and electrons must 
satisfy the single lepton selection criteria 
defined for the $4e$ and $4\mu$ final states.
In addition, electrons  and muons are
required to be spatially 
separated by $\Delta R > 0.2$ to remove $Z \to \mu\mu$ background
with muons radiating photons giving events with two muon and two 
trackless electron candidates. 
At least one muon must satisfy the same 
calorimeter isolation requirement 
imposed in the $4\mu$ final state.
A solution is required where one pair 
of same flavor leptons has an invariant mass $> 70$~GeV,
and the other $> 50$~GeV.
Finally, events are split
into three categories depending on the number of
electrons in the CC region. Subsamples with no electron,
with one electron and with two or more electrons in the CC are denoted
as $2\mu2e_{0C}$, $2\mu2e_{1C}$ and 
$2\mu2e_{2C}$, respectively.
As in the $4e$ channel, these subsamples have significantly 
different levels of background contamination.

A Monte Carlo (MC) simulation is used to determine 
the expected number of signal 
events in each subchannel. The small contribution from $ZZ$ events with
at least one $Z$ boson decaying into tau pairs is also included in the
signal. Simulated events are generated using 
\textsc{pythia}~\cite{pythia} and passed through
a detailed \textsc{geant}-based simulation~\cite{geant} of the detector
response. Differences
between MC simulation and data in the reconstruction and identification 
efficiencies for
electrons and muons are corrected using efficiencies derived from 
large data samples of inclusive $Z\rightarrow \ell\ell$ ($\ell = e$, or 
$\mu$) events.
The systematic uncertainty in the signal
is dominated by the uncertainty in the theoretical cross section
(6.25\%), the uncertainty on the lepton identification and reconstruction 
efficiencies 
($\approx4$\% for the $4e$ and $4\mu$ subchannels and $\approx2.5$\%
for the $2\mu2e$ subchannels) and a 6.1\% uncertainty on
the luminosity measurement~\cite{d0lumi}. Additional smaller sources of 
systematic uncertainty arise from energy and momentum resolutions and 
MC modeling of the signal process.  

Backgrounds to the $ZZ$ signal originate from top quark pair ($t\bar{t}$)
production and from events with $W$ and/or $Z$ bosons that decay to leptons 
and additional
jets or photons. The jets can then be misidentified as leptons or contain
true electrons or muons from in-flight decays of pions,
kaons, or heavy-flavored hadrons. 

The background from $t\bar{t}$ production is estimated from MC, 
assuming the cross section of 
$\sigma(t\bar{t})=7.9$~pb~\cite{ttbar_xsec} 
for a top quark mass of 170~GeV.
The systematic uncertainty includes a 10\% uncertainty on $\sigma(t\bar{t})$, 
as well as contributions from the variation in cross section
and acceptance originating from uncertainties on the mass of
the top quark.

To estimate the misidentified lepton background,
we first measure the probability for a jet to produce an 
electron or muon that satisfy the identification criteria from data
using a ``tag and probe'' method~\cite{runiia_zz}. 
The probability for a jet to mimic an electron, parameterized in
terms of jet $E_T$ and $\eta$, is equal to $4 \times 10^{-4}$ 
for the case of CC electrons with a 
matched track and $5 \times 10^{-3}$
in the case of EC electrons for which no track matching is applied. 
The probability for a 15~GeV (100~GeV) jet to produce a muon of 
$p_T>15$ GeV is $10^{-4} (10^{-2}) $ without requiring muon isolation,
and it is $10^{-5} (10^{-4})$
when the muon is required to be isolated.
A systematic uncertainty of 30\% on the jet-to-lepton misidentification
probabilities is estimated by varying the selection criteria of the
control samples used.

The probabilities for jets to be misidentified as
electrons are then applied to jets in $eee$+jets and
$\mu\mu e$+jets data to determine the background to the 
$4e$ and $2\mu2e$
channels, respectively. This method takes into account
contributions from $Z$+jets, $Z$+$\gamma$+jets,
$WZ$+jets, $WW$+jets, $W$+jets, and events with $\ge4$ jets. 
However, this method double counts the contribution from  
$Z$+jets. A correction is measured, 
amounting to $\approx20$\%, to 
correct for the double counting.
The probabilities for jets to contain a muon are applied to
jets in $\mu\mu$+jets data to determine a background estimate
for the $4\mu$ channel.
Systematic uncertainties on this background arise from the 30\% 
uncertainty in measured misidentification rates,
and from the limited statistics of the data remaining in the samples
after selection.

\begin{figure}
	\includegraphics[scale=0.40,angle=0]{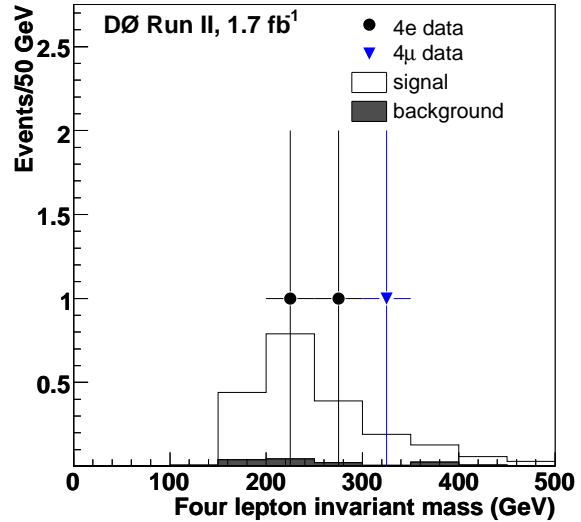}
	\caption{\label{fig:fourmass} Distribution of four lepton
	invariant mass in data, 
	expected signal, and expected background.}
\end{figure}

Table~\ref{tab:channels} summarizes the expected signal and background
contributions to each subchannel, as well as the number of candidate
events in data.
The total signal and background expectations are $1.89 \pm 0.08$
events and $0.14^{+0.03}_{-0.02}$ events, respectively.
We observe a total of three candidate events in the data, two
in the $4e_{4C}$ subchannel and one in the $4\mu$
subchannel.
Table~\ref{tab:candidates} summarizes some of their kinematic
characteristics.  
The quoted dilepton masses in the table correspond to the one out
of the three possible combinations having opposite charge within the
pairs and having a dilepton mass closest to that of the $Z$ boson.
Figure~\ref{fig:fourmass}
shows the distribution of the four lepton invariant mass 
for data and for the expected signal and
background.

\begin{table*}
\caption{The integrated luminosity, expected number of signal ($Z/\gamma^*$ 
$Z/\gamma^*$) and 
background events ($t\bar{t}$ and $Z(\gamma)$+jets which includes all 
$W$/$Z$/$\gamma$+jets
contributions), and the number of observed candidates in the 
seven $ZZ\to\ell^{+}\ell^{-}\ell^{'+}\ell^{'-}$ subchannels. Uncertainties
reflect statistical and systematic contributions added in quadrature.}
\vspace*{2mm}
\label{tab:channels}
\begin{tabular}{|c|c|c|c|c|c|c|c|} 
\hline \hline
Subchannel 
& $4e_{2C}$ 
& $4e_{3C}$ 
& $4e_{4C}$ 
& $4\mu$             
& $2\mu2e_{0C}$ 
& $2\mu2e_{1C}$ 
& $2\mu2e_{2C}$ \\ 
\hline
Luminosity (fb$^{-1}$) 
& $1.75 \pm 0.11$ 
& $1.75 \pm 0.11$ 
& $1.75 \pm 0.11$ 
& $1.68 \pm 0.10$ 
& $1.68 \pm 0.10$ 
& $1.68 \pm 0.10$ 
& $1.68 \pm 0.10$ \\ 
\hline 
 & & & & & & & \\[-2mm]
Signal 
& $0.084 \pm 0.008$ 
& $0.173 \pm 0.015$
& $0.140 \pm 0.012$ 
& $0.534 \pm 0.043$            
& $0.058^{+0.007}_{-0.006}$ 
& $0.352 \pm 0.040$ 
& $0.553^{+0.045}_{-0.044}$ \\[1mm] 
\hline
 & & & & & & & \\[-2mm]
$Z(\gamma)$+jets
& $0.030^{+0.009}_{-0.008}$ 
& $0.018^{+0.008}_{-0.007}$ 
& $0.002^{+0.002}_{-0.001}$ 
& $0.0003 \pm 0.0001$ 
& $0.03^{+0.02}_{-0.01}$ 
& $0.05 \pm 0.01$ 
& $0.008^{+0.004}_{-0.003}$ \\[1mm] 
\hline
 & & & & & & & \\[-2mm]
$t\bar{t}$ 
& -- 
& -- 
& -- 
& -- 
& $0.0012^{+0.0016}_{-0.0009}$ 
& $0.005 \pm 0.002$ 
& $0.0007^{+0.0009}_{-0.0005}$ \\[1mm] 
\hline
Observed events       
& 0       
& 0        
& 2          
& 1             
& 0       
& 0        
& 0 \\ 
\hline \hline
\end{tabular}

\end{table*}

\begin{table}
\caption{Characteristics of the observed candidate events. $\eta$ and
$\phi$ values are measured relative to the location of the 
$p\bar{p}$ collision. $M_{ll}$ is the mass of the lepton pair.}
\vspace*{2mm}
\label{tab:candidates}
  \begin{tabular}{|c|l|c|c|c|c|} 
    \hline \hline 
	&    
	& \ \ $e_1^+$ \ \ 
	& \ \ $e_2^+$ \ \ 
     	& \ \ $e_3^-$ \ \ 
	& \ \ $e_4^-$ \ \ \\
    	\cline{2-6}		
	& $p_T$ (GeV) 
	& 107 
	& 59 
	& 52 
	& 16 \\
     	\cline{2-6}
	$4e$
	& $\eta$ 
	& 0.66 
	& 0.25 
	& -0.64 
	& -0.85 \\
    	\cline{2-6}
	candidate 1  
	& $\phi$ 
	& 4.10
	& 1.08 
	& 0.46 
	& 2.62 \\
	\cline{2-6} 
	&   
	& \multicolumn{2}{c|}{$e_{1}^{+}e_{4}^{-}$} 
	& \multicolumn{2}{c|}{$e_{2}^{+}e_{3}^{-}$} \\
	& $M_{\ell\ell}$ (GeV) 
	& \multicolumn{2}{c|}{$89 \pm 3$} 
	& \multicolumn{2}{c|}{$61 \pm 2$} \\
    	\hline \hline
	&    
	& $e_1^+$ 
	& $e_2^+$ 
	& $e_3^-$ 
	& $e_4^-$ \\
    	\cline{2-6}
	& $p_T$ (GeV) 
	& 83 
	& 75 
	& 35 
	& 26 \\
    	\cline{2-6}
	$4e$
	& $\eta$ 
	& 0.64 
	& 0.40 
	& 0.85 
	& 1.17 \\
    	\cline{2-6}
	candidate 2  
	& $\phi$ 
	& 6.16
	& 3.80 
	& 3.83 
	& 1.40 \\
	\cline{2-6}
	&    
	& \multicolumn{2}{c|}{$e_{1}^{+}e_{3}^{-}$} 
	& \multicolumn{2}{c|}{$e_{2}^{+}e_{4}^{-}$} \\
	& $M_{\ell\ell}$ (GeV) 
	& \multicolumn{2}{c|}{$99 \pm 3$} 
	& \multicolumn{2}{c|}{$90 \pm 4$} \\
    	\hline \hline
	&        
	& $\mu_1^+$ 
	& $\mu_2^-$ 
	& $\mu_3^-$ 
	& $\mu_4^+$ \\
    	\cline{2-6}
	& $p_T$ (GeV) 
	& 115 
	& 77
	& 42 
	& 24 \\
    	\cline{2-6}
	$4\mu$
	& $\eta$ 
	& 0.04 
	& -1.01 
	& 0.77
	& -1.93 \\
    	\cline{2-6}
        candidate  
	& $\phi$ 
	& 1.69
	& 4.26 
	& 5.29 
	& 0.36 \\
	\cline{2-6}
	&    
	& \multicolumn{2}{c|}{$\mu_{1}^{+}\mu_{3}^{-}$} 
	& \multicolumn{2}{c|}{$\mu_{2}^{-}\mu_{4}^{+}$} \\[1mm]
	& $M_{\ell\ell}$ (GeV) 
	& \multicolumn{2}{c|}{$148^{+32}_{-18}$} 
	& \multicolumn{2}{c|}{$90^{+12}_{-8}$} \\[1mm]
    \hline \hline
  \end{tabular}
\end{table}

We extract the significance of the observed event distributions
using a negative log-likelihood ratio (LLR) 
test-statistic~\cite{collie}. As input, we use the expected
yields (number of events) from signal and background, separated into
the seven subchannels compared to the observed yields. 
The modified frequentist method returns 
the probability ($p$-value) of the background-only fluctuating
to give the observed yields or higher. In 5$\times10^9$
background pseudo-experiments, we find 213 trials with an LLR value
smaller or equal to that observed in data. This gives
a $p$-value of 4.3$\times$10$^{-8}$ which corresponds to a 5.3 
standard deviation ($\sigma$) observed significance ($3.7\sigma$ expected). 
The probability for the signal plus background hypothesis to give 
less signal-like observations than the observed one is 0.87. 
A correction factor of 0.93, derived using \textsc{pythia}, is used
to convert the measured cross section for $(Z/\gamma^{*})(Z/\gamma^{*})$
into that for $ZZ$ production.
Minimizing the LLR function we obtain a cross section of
$\sigma(ZZ) = 1.75^{+1.27}_{-0.86}~\mathrm{(stat.)} \pm 
0.13~\mathrm{(syst.)}$~pb for this
analysis.

This result is combined with the results from 
an independent $ZZ\to\ell^{+}\ell^{-}\nu\bar{\nu}$ 
search~\cite{llnunu}, and the previous  
$(Z/\gamma^{*})(Z/\gamma^{*}) \to \ell^{+}\ell^{-}\ell^{'+}\ell^{'-}$ 
analysis~\cite{runiia_zz} which
used a separate data sample with a looser mass requirement 
$M(\ell\ell) > 30$~GeV.
The earlier search contributes no signal events and we have scaled its
background estimate to the tighter kinematic range used in the recent
analysis. The combination of the three analyses is performed
taking into account the correlations of systematic uncertainties
between subchannels and among analyses.
The resulting $p$-value is 6.2$\times$10$^{-9}$, and 
the significance for observation of $ZZ$ production 
increases to $5.7\sigma$ ($4.8\sigma$ expected). The probability for 
the signal plus background hypothesis to give 
less signal-like observations than the observed one is 0.71. 
We therefore report the observation of a ZZ signal at a hadron collider with 
a combined cross section of $\sigma(ZZ) = 1.60 \pm
0.63~\mathrm{(stat.)}^{+0.16}_{-0.17}~\mathrm{(syst.)}$~pb, 
consistent with the standard model expectation.

%
We thank the staffs at Fermilab and collaborating institutions, 
and acknowledge support from the 
DOE and NSF (USA);
CEA and CNRS/IN2P3 (France);
FASI, Rosatom and RFBR (Russia);
CNPq, FAPERJ, FAPESP and FUNDUNESP (Brazil);
DAE and DST (India);
Colciencias (Colombia);
CONACyT (Mexico);
KRF and KOSEF (Korea);
CONICET and UBACyT (Argentina);
FOM (The Netherlands);
STFC (United Kingdom);
MSMT and GACR (Czech Republic);
CRC Program, CFI, NSERC and WestGrid Project (Canada);
BMBF and DFG (Germany);
SFI (Ireland);
The Swedish Research Council (Sweden);
CAS and CNSF (China);
Alexander von Humboldt Foundation (Germany);
and the Istituto Nazionale di Fisica Nucleare (Italy).
%


\begin{thebibliography}{99}
%
\bibitem[a]{alton}
Visitor from Augustana College, Sioux Falls, SD, USA.
\bibitem[b]{burdin}
Visitor from The University of Liverpool, Liverpool, UK.
\bibitem[c]{cerminara}
Visitor from INFN Torino, Torino, Italy.
\bibitem[d]{podesta-lerma}
Visitor from ECFM, Universidad Autonoma de Sinaloa, Culiac\'an, Mexico.
\bibitem[e]{quadt,meyer,hensel,park}
Visitor from II. Physikalisches Institut, Georg-August-University,
  G{\"o}ttingen, Germany.
\bibitem[f]{voutilainen}
Visitor from Helsinki Institute of Physics, Helsinki, Finland.
\bibitem[g]{weber}
Visitor from Universit{\"a}t Bern, Bern, Switzerland.
\bibitem[h]{wenger}
Visitor from Universit{\"a}t Z{\"u}rich, Z{\"u}rich, Switzerland.
\bibitem[\ddag]{deceased}
Deceased.

%
\vskip 0.25cm

\bibitem{brown} R.W. Brown and K.O. Mikaelian,
 Phys.\ Rev.\ D {\bf 19}, 922 (1979).

\bibitem{zz-theory} U. Baur and D. Rainwater,
 Phys.\ Rev.\ D {\bf 62}, 113011 (2000).

\bibitem{Campbell:1999ah}
  J.~M.~Campbell and R.~K.~Ellis,
  Phys.\ Rev.\  D {\bf 60}, 113006 (1999).

\bibitem{lep}
  R.~Barate {\it et al.}  [ALEPH Collaboration],
  Phys.\ Lett.\  B {\bf 469}, 287 (1999);
  J.~Abdallah {\it et al.}  [DELPHI Collaboration],
  Eur.\ Phys.\ J.\  C {\bf 30}, 447 (2003);
  M.~Acciarri {\it et al.}  [L3 Collaboration],
  Phys.\ Lett.\  B {\bf 465}, 363 (1999);
  G.~Abbiendi {\it et al.}  [OPAL Collaboration],
  Eur.\ Phys.\ J.\  C {\bf 32}, 303 (2003).

\bibitem{runiia_zz}
  V.~M.~Abazov {\it et al.}  [D0 Collaboration],
  Phys.\ Rev.\ Lett.\  {\bf 100}, 131801 (2008).

\bibitem{cdf_zz}
  T.~Aaltonen {\it et al.}  [CDF Collaboration],
  Phys.\ Rev.\ Lett.\  {\bf 100}, 201801 (2008).

\bibitem{d0lumi}
  T.~Andeen {\it et al.} , FERMILAB-TM-2365 (2007).

\bibitem{run2det}
  V.~M.~Abazov {\it et al.}  [D0 Collaboration],
  Nucl.\ Instrum.\ Meth. Phys. Res.\  A {\bf 565}, 463 (2006).

\bibitem{pseudo}
     The D0  coordinate system is cylindrical with the 
     $z$-axis along the proton beamline and the polar and azimuthal
     angles denoted as $\theta$ and $\phi$ respectively. The 
     pseudorapidity is defined as $\eta=-\ln[\tan(\theta /2)]$.

\bibitem{cal_iso} The variable $\Delta R$ between two objects
$i$ and $j$ is defined as $\Delta R = \sqrt{(\eta_{i}-\eta_{j})^{2} +
(\phi_{i}-\phi_{j})^{2}}.$

\bibitem{pythia}
  T.~Sj\"{o}strand {\it et al.},
  Comput.\ Phys.\ Commun.\  {\bf 135}, 238 (2001).

\bibitem{geant}
  R. Brun and F. Carminati, CERN Program Library Long Writeup W5013, 
  1993 (unpublished).

\bibitem{ttbar_xsec}
  N.~Kidonakis and R.~Vogt,
  Phys.\ Rev.\  D {\bf 68}, 114014 (2003).

\bibitem{collie} W.~Fisher, FERMILAB-TM-2386-E (2007).

\bibitem{llnunu}
  V.~M.~Abazov {\it et al.}  (D0 Collaboration),
  Phys.\ Rev.\ D {\bf 78}, 072002 (2008).

\end{thebibliography}
\end{document}